\documentclass[aps,twocolumn,pra,groupedaddress,showpacs]{revtex4-1}

\usepackage{graphicx}

\begin{document}


\title{Photo induced ionization dynamics of the nitrogen vacancy defect in diamond investigated by single shot charge state detection}


\author{N.~Aslam$^1$}
\email{n.aslam@physik.uni-stuttgart.de}
\author{G.~Waldherr$^1$}
\author{P.~Neumann$^1$}
\author{F.~Jelezko$^2$}
\author{J.~Wrachtrup$^1$}
\affiliation{$^1$3. Physikalisches Institut and Research Center SCOPE, University of Stuttgart, 70569 Stuttgart, Germany}
\affiliation{$^2$ Institut f\"{u}r Quantenoptik, Universit\"{a}t Ulm, Ulm D-89073, Germany}


\date{\today}

\begin{abstract}
The nitrogen-vacancy centre (NV) has drawn much attention for over a decade, yet detailed knowledge of the photophysics needs to be established.
Under typical conditions, the NV can have two stable charge states, negative (NV$^-$) or neutral (NV$^0$), with photo induced interconversion of these two states.
Here, we present detailed studies of the ionization dynamics of single NV centres in bulk diamond at room temperature during illumination in dependence of the excitation wavelength and power.
We apply a recent method which allows us to directly measure the charge state of a single NV centre, and observe its temporal evolution.
Results of this work are the steady state NV$^-$ population, which was found to be always $\leq 75$\% for 450 to 610~nm excitation wavelength, the relative absorption cross-section of NV$^-$ for 540 to 610~nm, and the energy of the NV$^-$ ground state of 2.6~eV below the conduction band.
These results will help to further understand the photo-physics of the NV centre.
\end{abstract}


\maketitle

\section{Introduction}
\begin{figure}
	\centering
		\includegraphics[width=0.48\textwidth]{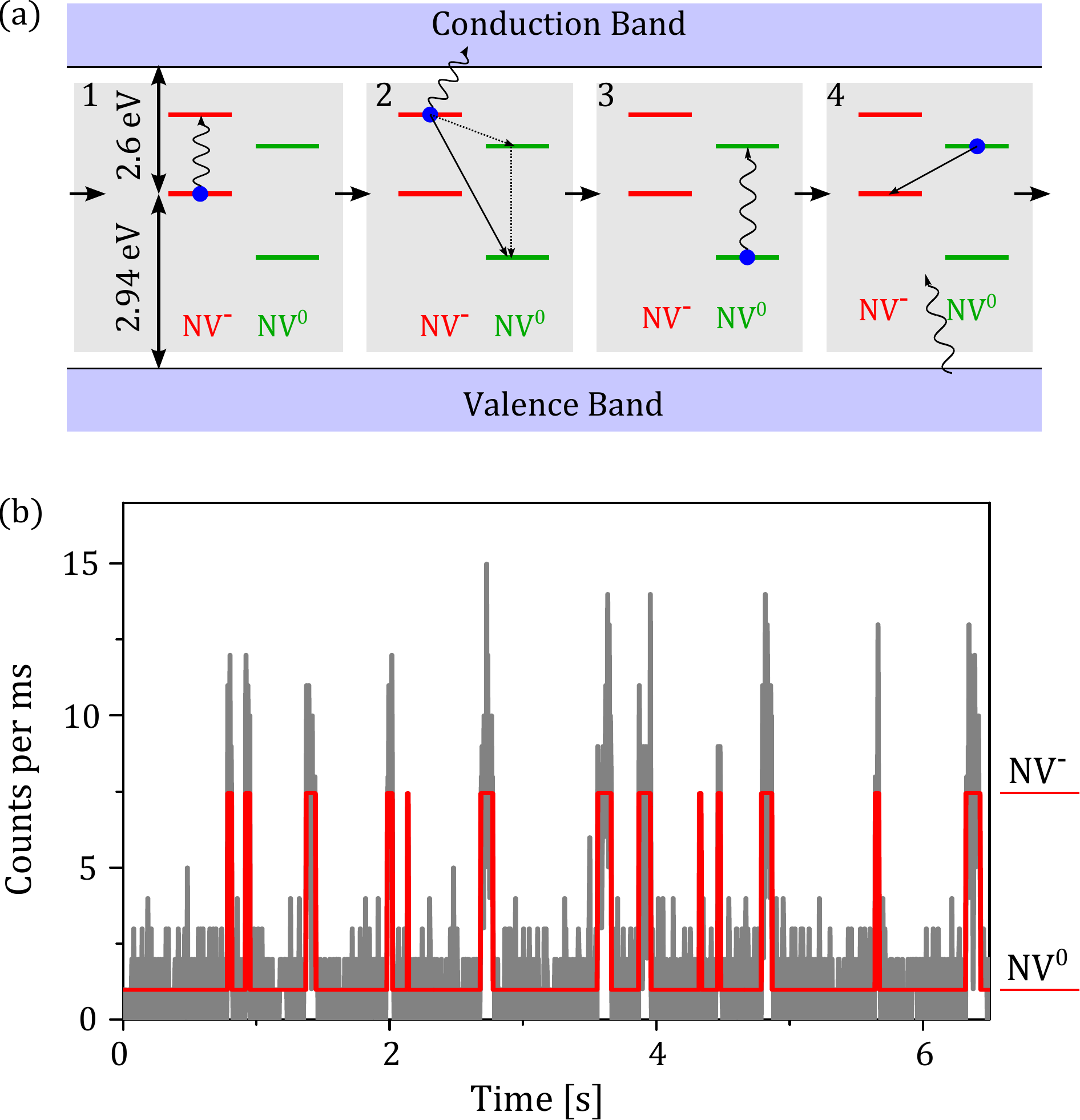}
	\caption{(a) Schematic illustration of the photo induced ionization and recombination of NV$^-$, showing the ground and excited states of both NV$^-$ and NV$^0$.
	The determination of the ionization energy of 2.6~eV and the recombination energy of 2.94~eV is shown in Fig.~\ref{fig:Fig_Ionization_Energy}.
	Ionization process:
	1. NV$^-$ is excited by one photon.
	2. A second photon excites one electron from the excited state of NV$^-$ into the conduction band, creating NV$^0$.
	The NV will end up in the NV$^0$ ground state, either directly after the ionization, or possibly via the excited state and subsequent, fast decay.
	Recombination process:
	3. NV$^0$ is excited.
	4. An electron is captured from the valence band, forming NV$^-$.
	(b) Timetrace of the fluorescence of a single NV under continuous illumination with 593~nm, 1~$\mu$W laser light, showing a dark state of the NV (which turns out to be NV$^0$).
	The red line is the most likely fluorescence level of the NV, obtained by a hidden Markov model.
	Here, the lifetimes of the two charge states are $T_{NV^-}=56.6$~ms, $T_{NV^0}=465$~ms, and the fluorescence levels are $Fl_{NV^-}=2.2$ counts/ms, $Fl_{NV^0}=0.3$ counts/ms.}
	\label{fig:Fig_energy_levels}
\end{figure}

The interesting features of the negatively charged NV$^-$ centre are high and stable fluorescence, optical spin state detection, long spin coherence times, and coherent control of surrounding nuclear spins via hyperfine interaction \cite{Kurtsiefer_PRL_2000, Gruber_Science1997, Jelezko_PRL_20041, Jelezko_PRL_20042, Balasubramanian_NatMater_2009}.
Promising applications are the use as fluorescence markers \cite{Han_NanoLetters_2010, Tisler_ACSNano_2011, McGuinness_NatNano_2011}, nanoscale magnetic and electric field sensing \cite{Balasubramanian_Nature_2008, Maze_Nature_2008, Tayl_natphys_2008, Dolde_nphys_2011}, quantum information processing \cite{Neumann_Science_2010, Dutt_Science_2007, Togan_nature_2010, Robledo_nature_2011}, and fundamental studies of spin bath physics \cite{Hanson_Science_2008, Lange_Science_2010}.
For all these applications, understanding and control of the NV centre charge dynamics is important.
\\
\indent
The NV consists of a substitutional nitrogen atom and an adjacent vacancy in the diamond carbon lattice.
Its electronic structure is built from the three carbon dangling bond electrons, two electrons from the nitrogen lone pair orbital for the neutral charge state NV$^0$, and one more electron for the negative charge state NV$^-$.
The zero-phonon line of NV$^-$ is at 637~nm, and phonon assisted sidebands of the fluorescence extend up to 800~nm.
The spectrum of NV$^0$ has a zero-phonon line at 575~nm and also exhibits strong Stokes shifted vibronic bands.
Both charge states can be observed in diamond.
Our current understanding of the photochromism of the NV is schematically shown in Fig.~\ref{fig:Fig_energy_levels}(a) \cite{Waldherr_PRL_20111, Siyushev_arxiv_2012}.
Figure~\ref{fig:Fig_energy_levels}(b) shows real-time monitoring of the charge state, which we use to analyse the ionization dynamics \cite{Waldherr_PRL_20112}.
\\
\indent
So far, the wavelength and power dependence of the photo induced ionization and recombination of NV$^-$ has only been investigated partly \cite{Mans_Diam_2005, Iak_JPCondM_2000, Kupr_JPCondM_2000, Steeds_diamond_2000, Gaebel_APB_2006, Han_NanoLetters_2010, Waldherr_PRL_20111, Waldherr_PRL_20112, Siyushev_arxiv_2012}.
Pumping NV$^-$ into an unknown dark state under red illumination was used for high resolution microscopy in \cite{Han_NanoLetters_2010}.
By single shot NMR measurements on single NV$^-$, it was found that under typically used 532~nm illumination, NV$^-$ is with 30 \% probability in this dark state, which was proposed to be NV$^0$ \cite{Waldherr_PRL_20111}.
Here, we will provide further evidence that the unknown dark state is indeed the neutral charge state NV$^0$.
\\
\indent
The experiments were performed using a homebuilt confocal microscope, and measuring the NV fluorescence either by an APD or a spectrometer.
We used a pulsed supercontinuum laser (SC400-4-PP system from Fianium) with 40~MHz repetition rate combined with an acousto optic tunable filter (AOTF) to select the desired wavelength.
The resulting spectrum had a width of about $4\;$nm around the selected wavelength.
Additionally, a $532\;$nm and a 594~nm DPSS lasers were used.
All lasers were focused onto a diffraction limited spot.
The used samples were ultrapure IIa type chemical vapour deposition (CVD) diamonds.
For our measurements we use single shot NMR \cite{Neumann_Science_2010} and single shot charge state detection \cite{Waldherr_PRL_20112}.
\\
\indent
The paper is organized as follows:
In Sec.~\ref{sec:dark_state}, we show that the above mentioned dark state of NV$^-$ is indeed the neutral NV$^0$, as recently proposed.
In Sec.~\ref{sec:ion_rates}, we investigate the photo-induced ionization and recombination dynamics in dependence of excitation wavelength and power.
With this, we can determine the excitation spectrum of NV$^-$ and the steady state population of the two charge states.
In Sec.~\ref{sec:cds_saturation}, the steady state population of NV$^-$ is measured over a broad spectral range of the excitation wavelength.
Together with the results from Sec.~\ref{sec:ion_rates}, we find that the population is always $\leq$ 75\%.
Additionally, we determine the fluorescence saturation behaviour of NV$^-$, which, combined with the population of NV$^-$, yields the optimal excitation wavelength.
This is found to be around 510 to 540~nm.
Finally, in Sec.~\ref{sec:ion_energy}, the ionization and recombination energies of NV$^-$ are determined.
\\
\section{NV$^-$ "Dark State" is NV$^0$}\label{sec:dark_state}
\indent
\begin{figure}
	\centering
		\includegraphics[width=0.48\textwidth]{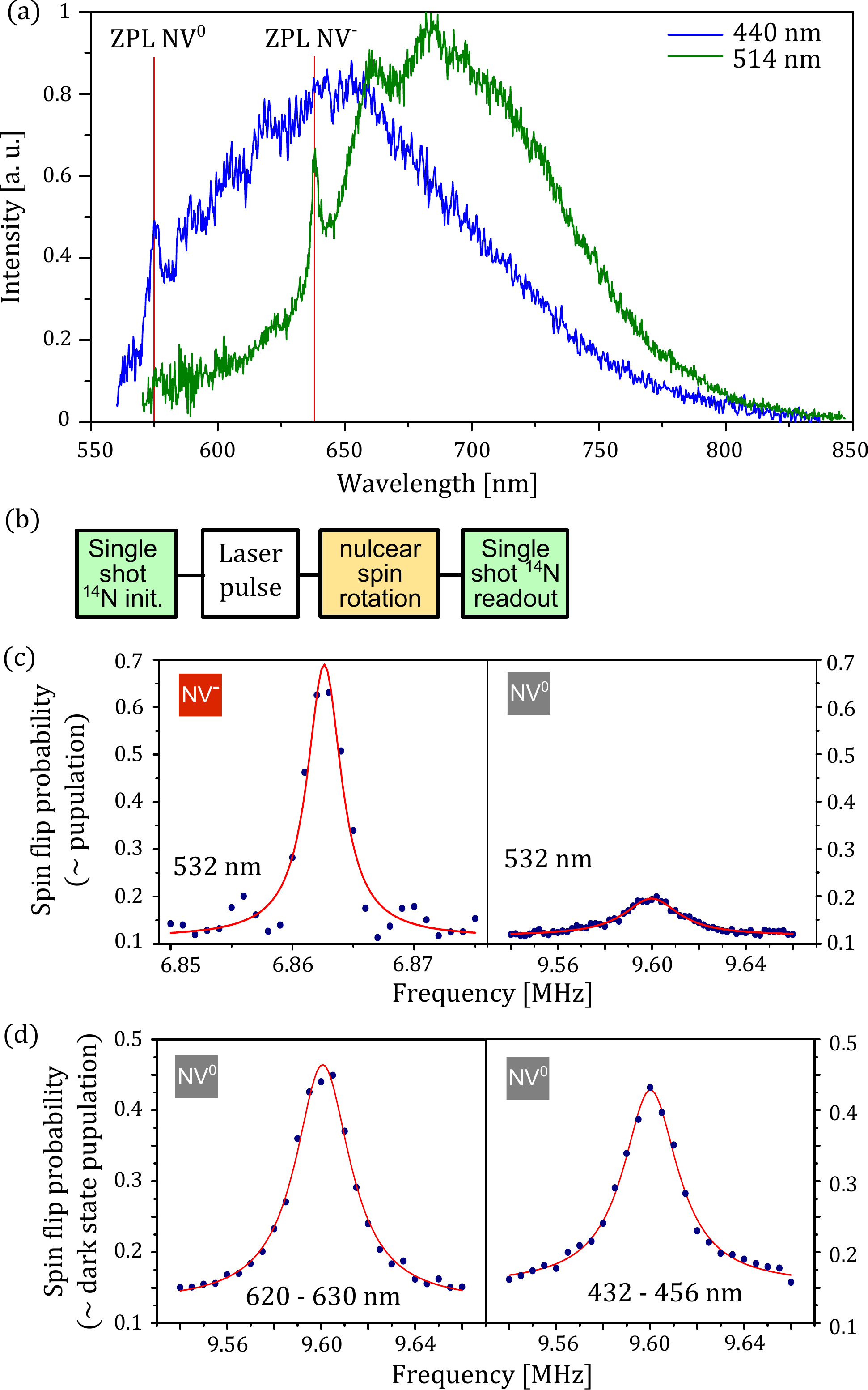}
	\caption{(a) Emission spectra of a single NV by illumination with 440~nm (yielding the spectrum of NV$^0$) and 514~nm (yielding the spectrum of NV$^-$). 
	Both spectra are normalized to have the same area.
	(b) Measurement sequence for (c).
	First, a single shot measurement determines the state of the nitrogen nuclear spin.
	Then, a variable laser pulse is applied.
	The nuclear spin state is very robust with respect to excitation and ionization of the NV, and survives these processes.
	The following RF pulses flips the nuclear spin, if it is in resonance to a nuclear spin transition, which depends on the electronic spin state.
	Finally, the state of the nuclear spin is measured.
	Comparing the initial and final measurement yields the probability to flip the nuclear spin.
	Further details can be found in \cite{Waldherr_PRL_20111}.
	(c) NMR spectra for green (532~nm) illumination, for a transition ($m_{\mathrm{N}} +1 \leftrightarrow 0$) in the NV$^-$ ground state (left) and in the unknown dark state (right).
	Here, no extra laser pulse with 532~nm was applied, because the nuclear spin state measurement includes these pulses.
	(d) NMR spectra for red (625~nm) and blue (440~nm) illumination.
	For 620 - 630~nm, the laser power was 20~$\mu$W and the pulse length 2~ms.
	For 432 - 456~nm, it was 20~$\mu$W for 10~ms.
	The red lines are Lorentzian fits.
	Note that for this transition, there is fast homogeneous dephasing, and at most 0.5 spin flip probability can be achieved for 100 \% population of the corresponding electronic state \cite{Waldherr_PRL_20111}.}
	\label{fig:Fig_DarkStateNMR}
\end{figure}
By illuminating NV with light of different wavelengths and measuring the fluorescence spectrum, we find that excitation with blue light (440~nm) populates the neutral charge state NV$^0$.
At these wavelengths, the emission of NV$^0$ is increased, whereas nearly no emission of NV$^-$ is observable, see Fig.~\ref{fig:Fig_DarkStateNMR}(a).
As explained in \cite{Waldherr_PRL_20111}, we can use single shot NMR (see \cite{Neumann_Science_2010}) on the NV's nitrogen nuclear spin to probe the population of an electronic state of the NV.
This is because the energy splitting of the nuclear spin depends on the hyperfine interaction to the NV electron spin and the quadrupole interaction of the nucleus, which both depend on the electronic wave function being different for NV$^-$ and NV$^0$. 
The probability to flip the nuclear spin at a transition frequency corresponding to a certain electronic state depends on the population of this state.
The NMR measurement sequence is shown in Fig.~\ref{fig:Fig_DarkStateNMR}(b).
Figure \ref{fig:Fig_DarkStateNMR}(c) shows NMR spectra for the $^{14}$N nuclear spin transition in the NV$^-$ ground state and unknown dark state after green (532~nm) illumination, with different transition frequencies.
We then apply blue (440~nm) and red (625~nm) illumination before the NMR pulse, and afterwards measure the probability that the nuclear spin is flipped at the transition frequency corresponding to the dark state. 
The resulting NMR spectra are shown in Fig.~\ref{fig:Fig_DarkStateNMR}(d).
We find that both red and blue illumination populate this state.
Because blue light populates NV$^0$ (see above), the unknown dark state is NV$^0$.
\\
\section{Ionization Dynamics}\label{sec:ion_rates}
\indent
To gain further insight into the photo induced ionization and recombination of NV$^-$, we perform real-time monitoring of the NV charge state by measuring NV fluorescence \cite{Waldherr_PRL_20112}.
We use a 650~nm long pass filter to mainly detect emission from NV$^-$.
Especially for photon energies less than the zero phonon line of NV$^0$ (575~nm), NV$^0$ cannot be excited, and will therefore not fluoresce.
Therefore, NV$^-$ will give a high fluorescence signal, whereas NV$^0$ is dark.
An example time trace of the NV$^-$ fluorescence with 593~nm, 1~$\mu$W illumination is shown in Fig.~\ref{fig:Fig_energy_levels}(b).
This real-time monitoring of the NV charge state is only possible if the photon count difference from NV$^-$ and NV$^0$ is larger than the photon shot noise, which is the case for low power illumination:
The ionization rate $r_{\rm{Ion}}$ and recombination rate $r_{\rm{Re}}$ both depend quadratically on the excitation power $p$, i.e. $r_{\rm{Ion}}, r_{\rm{Re}} \propto p^2$.
The NV$^-$ fluorescence $fl_{NV^-}$, on the other hand, depends linearly on the power, $fl_{NV^-} \propto p$.
Effectively, the average number of photons we detect from NV$^-$ before it is ionized is $photons \propto fl_{NV^-}/r_{\rm{Ion}} \propto \frac{1}{p}$, i.e. it increases linearly with decreasing power.
To analyse the ionization dynamics, we use a hidden Markov model with two possible fluorescence levels, which yields the most likely state evolution, and also the corresponding lifetimes and transition rates \cite{Zarrabi_SPIE_2008}.
\\
\subsection{Power Dependence}
\indent
\begin{figure}
	\centering
		\includegraphics[width=0.48\textwidth]{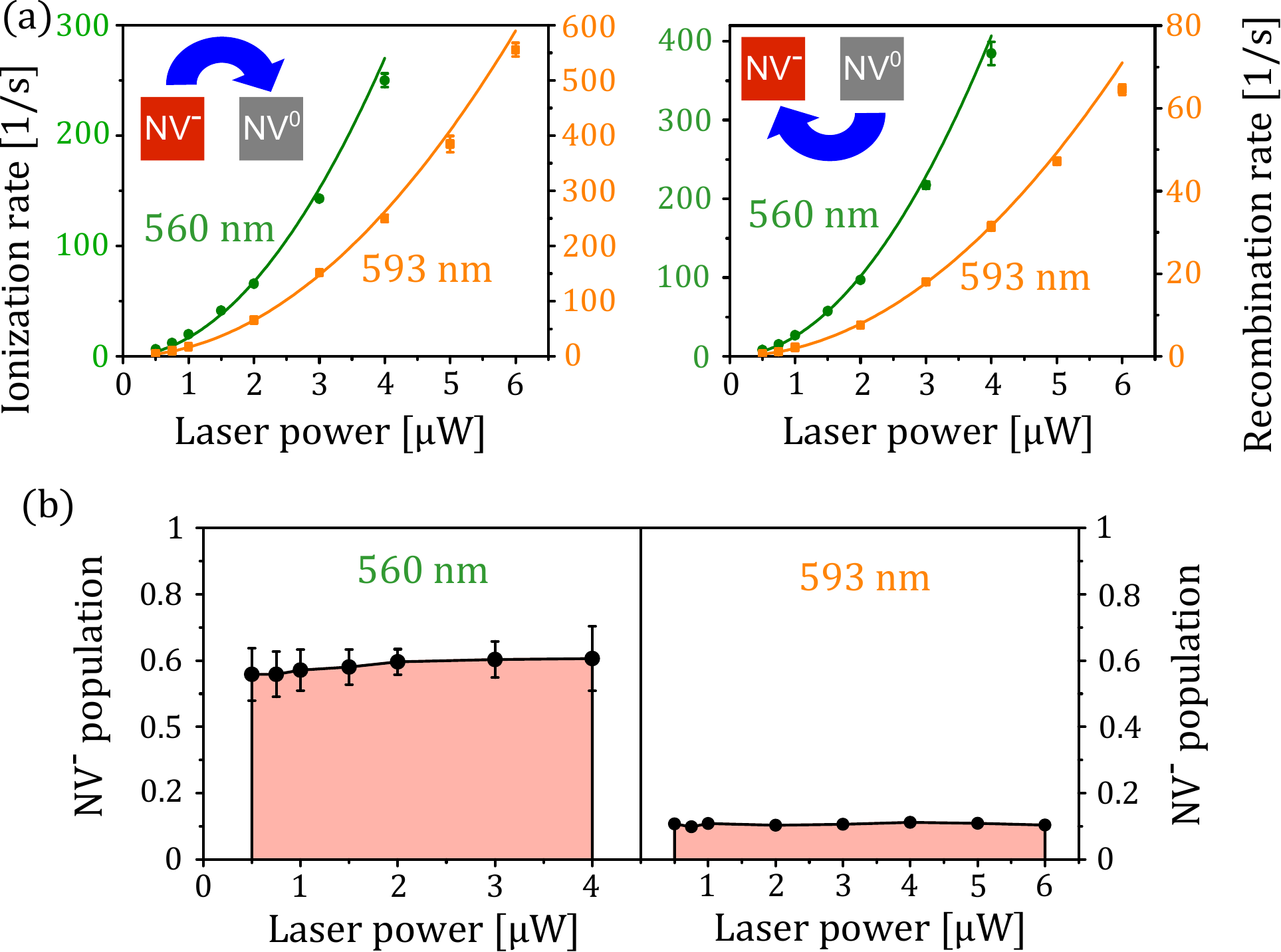}
	\caption{a) Power dependence of the ionization and recombination rates for 560~nm and 593~nm illumination wavelength.
	The solid lines are fits with quadratic power dependence, $\mathrm{rate}(p) = ap^2$, where $p$ is the laser power.
	b) Probability to be in NV$^-$ in dependence of the illumination power for 560~nm and 593~nm.}
	\label{fig:Fig_Ion_rates_power}
\end{figure}
First, we performed real-time monitoring of the charge state (cf. Fig.~\ref{fig:Fig_energy_levels}(b)) for varying excitation power from 0.5 - 6~$\mu$W for 560~nm and 593~nm excitation wavelength.
Figure~\ref{fig:Fig_Ion_rates_power}(a) shows the ionization and recombination rates for these illumination parameters.
The curves show the expected quadratic behaviour for low excitation power for both wavelengths.
In Fig.~\ref{fig:Fig_Ion_rates_power}(b), we calculate the population of NV$^-$ in dependence of the power by detailed balance as
\begin{equation}\label{eq:pnv-}
P_{\rm{NV}^-} = \frac{T_{\rm{NV}^-}}{T_{\rm{NV}^-}+T_{\rm{NV}^0}},
\end{equation}
where $T_{\rm{NV}^-}$ and $T_{\rm{NV}^0}$ are the charge state lifetimes of NV$^-$ and NV$^0$ obtained from the fluorescence time trace (here, the lifetime is the inverse of the transition rate).
The population is nearly independent of the power, because both rates (ionization and recombination) have a quadratic dependence.
\subsection{Wavelength Dependence}
\indent
\begin{figure}
	\centering
		\includegraphics[width=0.48\textwidth]{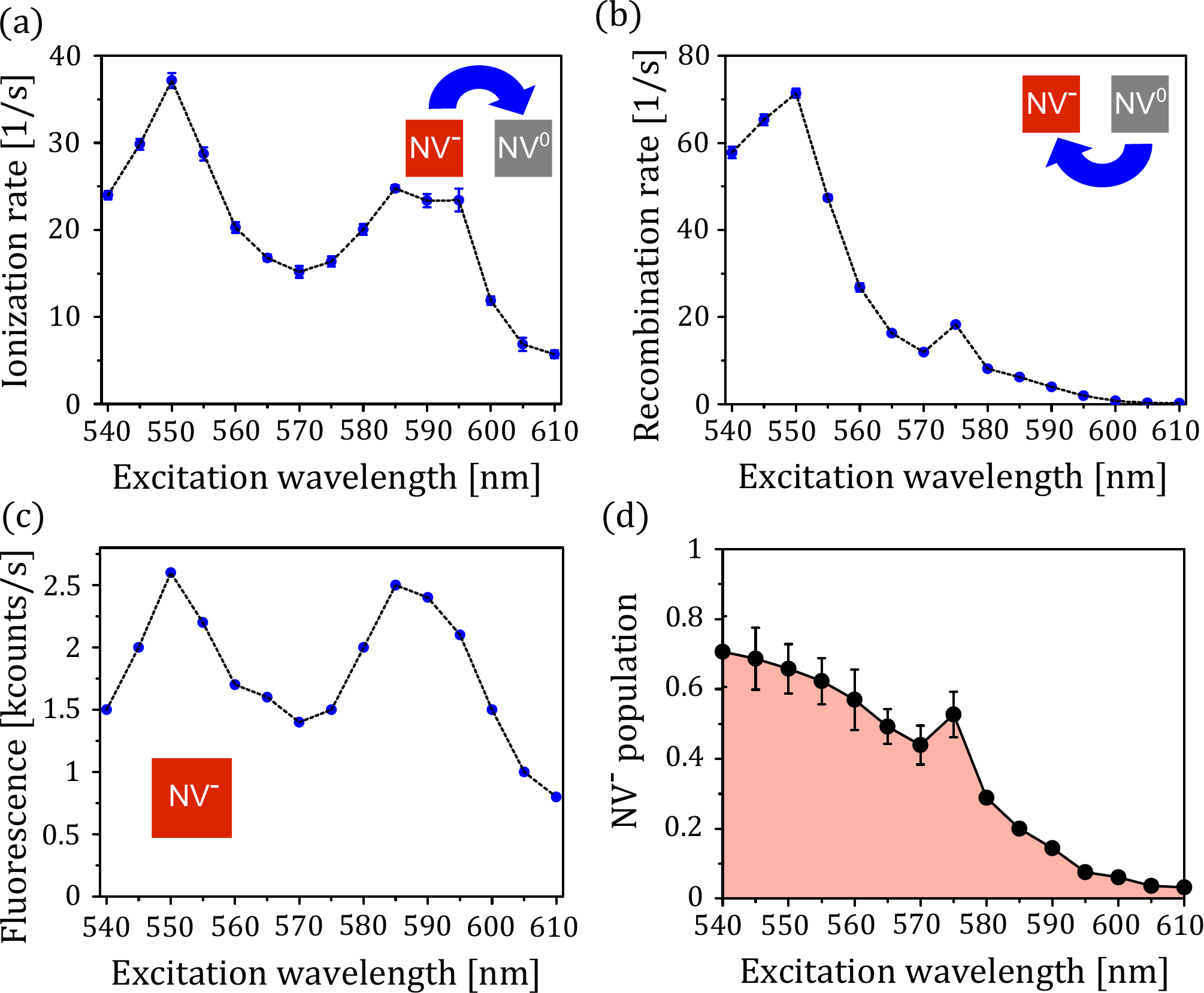}
	\caption{Results obtained from the ionization dynamics in dependence of the excitation wavelength at a fixed power of 1~$\mu$W.
	a) Ionization and b) recombination rates.
	c) Fluorescence of NV$^-$. Because the power is fixed, this is proportional to the absorption cross section of NV$^-$.
	d) Probability to be in NV$^-$ during illumination.
	The local maximum at 575~nm is at the zero phonon line of NV$^0$.}
	\label{fig:Fig_Ion_rates_wavelength}
\end{figure}
Real-time monitoring of the charge state for varying excitation wavelength from 540 - 610~nm (fixed excitation power of 1~$\mu$W) yields the \textit{bare} absorption cross-section of NV$^-$ and the population ratio of NV$^-$ to NV$^0$.
For wavelengths shorter than 540~nm fluorescence of NV$^-$ and NV$^0$ cannot be resolved, because the fluorescence contrast as well as the charge state lifetimes decrease.
For wavelengths longer than 610~nm, the recombination rate from NV$^0$ to NV$^-$ becomes too small. 
Figures~\ref{fig:Fig_Ion_rates_wavelength}(a) and (b) show the ionization and recombination rates in dependence of the illumination wavelength.
For the recombination rate shown in Fig.~\ref{fig:Fig_Ion_rates_wavelength}(b), we find a local maximum at 575~nm, which is the zero phonon line of NV$^0$.
This is expected, as recent low-temperature studies have shown that for the recombination process, first NV$^0$ has to be excited, and then an electron from the valence band is captured \cite{Siyushev_arxiv_2012}, see fig.~\ref{fig:Fig_energy_levels}(a).
For shorter wavelengths, the recombination rate increases, probably due to increased absorption of NV$^0$ into the phononic sideband of its excited state, with a maximum at 550~nm.
The ionization rate in Fig.~\ref{fig:Fig_Ion_rates_wavelength}(a) shows two maxima at 550~nm and 590~nm.
As explained in the following, this results mainly from the absorption cross-section of NV$^-$, i.e. due to the increased excitation rate of NV$^-$ at these wavelengths.
In the example fluorescence time trace shown in Fig.~\ref{fig:Fig_energy_levels}(b), high fluorescence indicates emission from NV$^-$ only.
This means that the obtained fluorescence level is directly proportional to the excitation efficiency (absorption cross-section) of NV$^-$.
In contrast to this, for high power illumination the real-time monitoring of the charge state is not possible, and observed fluorescence of the NV would be a mixture of the two charge states NV$^-$ and NV$^0$.
The fluorescence excitation efficiency of NV$^-$ in Fig.~\ref{fig:Fig_Ion_rates_wavelength}(c) also shows two maxima at 550~nm and 585-590~nm, similar to the ionization rate.
Note that the curve does not resemble the mirrored fluorescence spectrum of NV$^-$ shown in Fig.~\ref{fig:Fig_DarkStateNMR}(a).
The double peak structure could result from differing vibrational modes in the ground and excited states, or possibly from a second electronically excited state involved in the excitation process \cite{Band_PRA_1988}.
\\
\indent
Another important point is the population of NV$^-$ during illumination, because this is the desired charge state for most applications.
This can be calculates according to eq.~(\ref{eq:pnv-}).
The result is shown in Fig.~\ref{fig:Fig_Ion_rates_wavelength}(d).
We can see that $P_{\rm{NV}^-}$ decreases for increasing wavelength above 540~nm, with a local maximum at 575~nm due to the decreased lifetime of NV$^0$.
Populating NV$^-$ with $> 75\%$ probability cannot be achieved at these wavelengths.
\\
\indent
\\
\section{NV$^-$ Population and Saturation Behaviour}\label{sec:cds_saturation}
\begin{figure}
	\centering
		\includegraphics[width=0.42\textwidth]{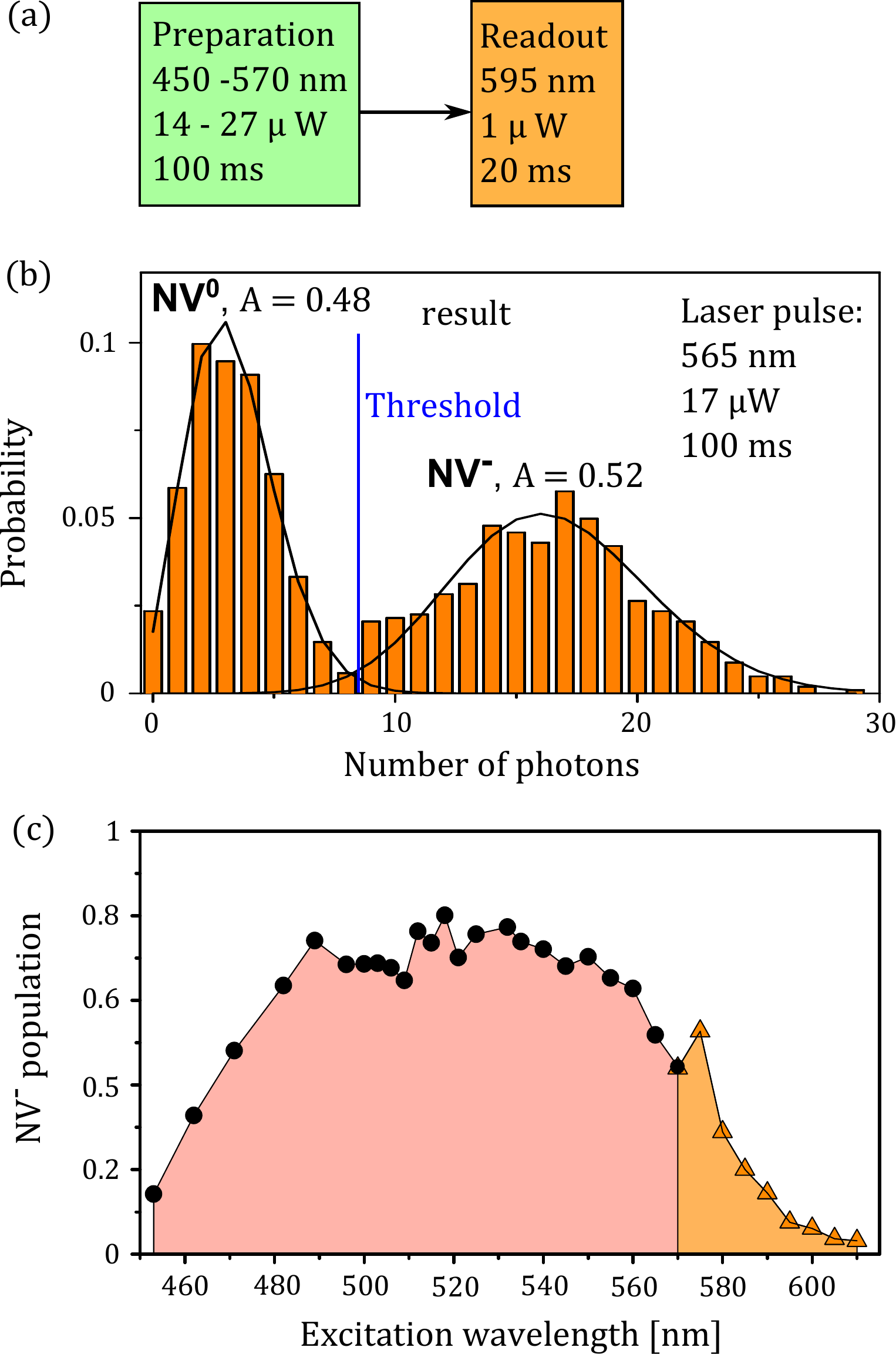}
	\caption{a) Pulse sequence for measuring the probability to be in NV$^-$ for arbitrary illumination.
	The first laser pulse populates NV$^-$ and NV$^0$ with a certain probability, which is measured by the second pulse.
	A wavelength of 594~nm is optimum for the measurement pulse, because absorption of NV$^-$ is high [cf. Fig.~\ref{fig:Fig_Ion_rates_wavelength}(c)], leading to high NV$^-$ fluorescence.
	b) Histogram of photon counts during the 594~nm measurement pulse, after illuminating the NV with 565~nm, 22.5~$\mu$W for 100~ms.
	The solid lines are obtained from fitting the sum of two Poissonian distributions.
	The amplitudes of the two Poissonians is indicated by $A$.
	The blue line indicates a threshold, which can be used to determine the charge state of the NV after one single measurement.
	c) Steady state probability to be in NV$^-$ in dependence of the illumination wavelength, obtained from the amplitudes of the Poissonian fits (black dots).
	The data above 570~nm (orange triangles) is taken from Fig.~\ref{fig:Fig_Ion_rates_wavelength}(d).}
	\label{fig:Fig_charge_state_detection}
\end{figure}

\begin{figure}
\centering
		\includegraphics[width=0.45\textwidth]{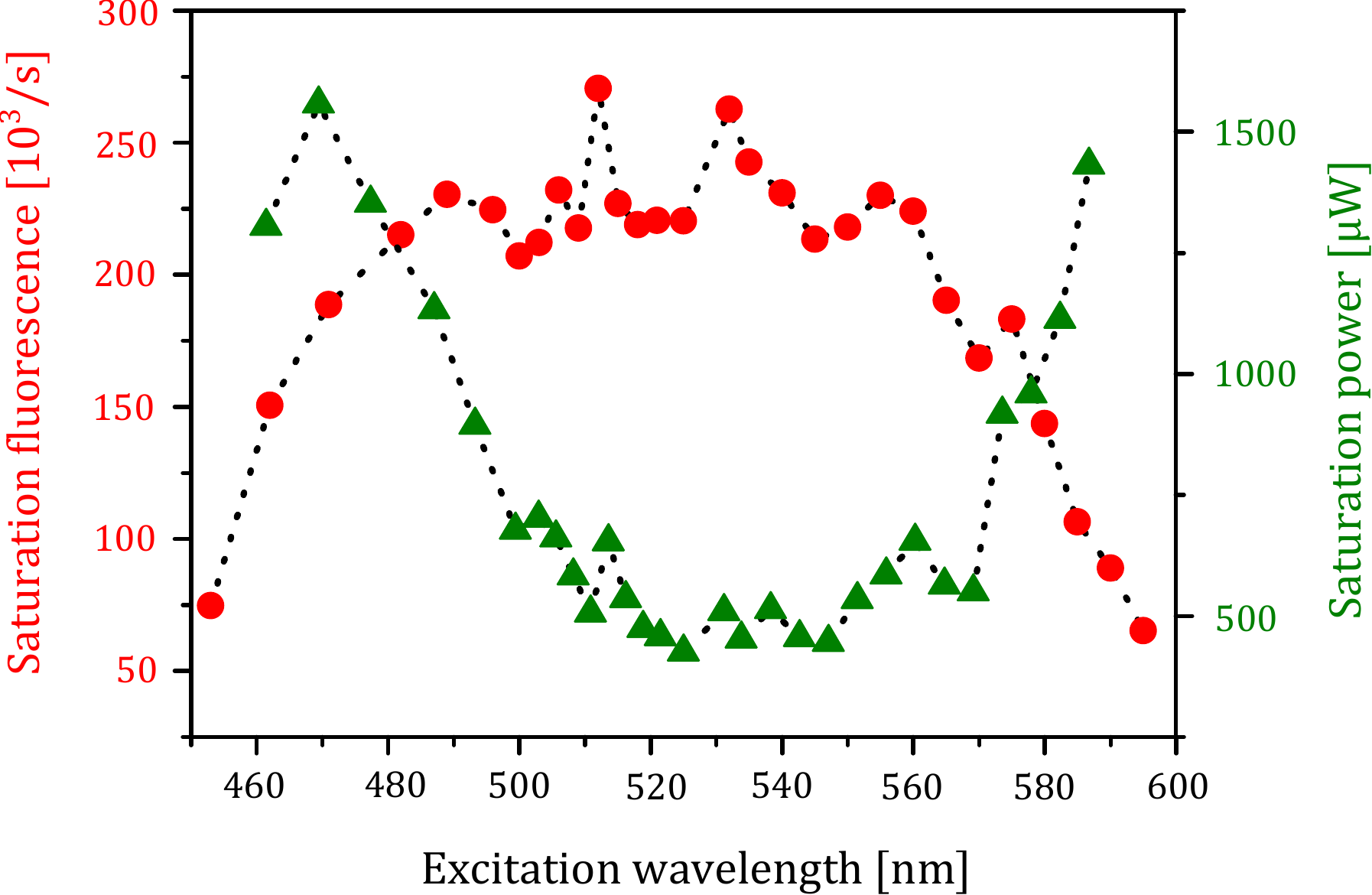}
	\caption{Saturation fluorescence (red dots) and saturation power (green triangles) of the NV, obtained from saturation curves.
	The peaks at around 513 and 530~nm are probably errors, and could not be confirmed by separate measurements.
	The local peak at 575~nm, however, is due to the zero phonon line of NV$^0$, and can also be seen in Fig.~\ref{fig:Fig_charge_state_detection}(c)}
	\label{fig:Fig_Saturation}
\end{figure}

\indent
To obtain the population of NV$^-$ for shorter wavelengths than 540 nm, we use a single shot charge state detection method \cite{Waldherr_PRL_20112}.
As explained above and shown in Fig.~\ref{fig:Fig_energy_levels}(b), low power illumination allows direct observation of the NV's charge state.
By properly choosing the wavelength, power and duration of a single laser pulse, the number of fluorescence photons counted during this pulse allows single shot determination of the charge state.
The histogram of many such measurements shows two Poisson distributions corresponding to the fluorescence of NV$^0$ and NV$^-$, see Fig.~\ref{fig:Fig_charge_state_detection} b).
By applying an arbitrary laser pulse before performing this charge state measurement, we can measure the probability to be in NV$^-$ after this laser pulse, as shown in Fig.~\ref{fig:Fig_charge_state_detection}(a).
By fitting two Poisson distributions to the histogram, the ratio of the amplitudes yields the probability $P_{\rm{NV}^-}$ to be in NV$^-$:
\begin{equation}
P_{\rm{NV}^-} = \frac{A_{\rm{NV}^-}}{A_{\rm{NV}^-}+A_{\rm{NV}^0}},
\end{equation}
where $A_{\rm{NV}^-}$ and $A_{\rm{NV}^0}$ are the amplitudes of the Poissonian fit functions.
Note that the first laser pulse in sequence Fig.~\ref{fig:Fig_charge_state_detection}(a) must be long enough to reach a steady state population of the two charge states.
\\
\indent
The result is shown in Fig.~\ref{fig:Fig_charge_state_detection}(c).
For 540~nm to 570~nm, we obtain similar results as above (Fig.~\ref{fig:Fig_Ion_rates_wavelength}).
Between 510~nm to 540~nm, the probability to be in NV$^-$ is around 75\%, whereas for $<$ 500~nm the probability decreases.
This means that there is no particular wavelength at which the NV can be initialized into NV$^-$ with $> 75\%$ probability.
\\
\indent
Another interesting point is the saturation behaviour of the NV under high power illumination, in dependence of the illumination wavelength.
On the one hand, this yields the maximal fluorescence (saturation fluorescence) of the NV, which should be proportional to the probability to be in NV$^-$.
On the other hand, it also yields the saturation power, i.e. how much power is needed to get half of the maximum fluorescence.
The desired condition for experiments is a high maximal fluorescence combined with low saturation power.
For this measurement, we collected saturation curves by determining the fluorescence in dependence of the illumination power.
The function used to fit the measured data and get the saturation fluorescence $F_S$ and the saturation power $I_S$ is
\begin{equation}\label{eq:saturation}
F = F_S\frac{I}{I+I_S},
\end{equation}
where F is the fluorescence and I the illumination power (see Appendix~\ref{app:saturation}).
For moderate powers we are using ($<$ 1 mW), this function fits the data very well.
The result of this measurement is shown in Fig.~\ref{fig:Fig_Saturation}.
For illumination with 520 to 550~nm, we get high saturation fluorescence and low saturation power.
As expected, the shape of the saturation fluorescence looks very similar to the population of NV$^-$ at low powers shown in Fig.~\ref{fig:Fig_charge_state_detection}(c).
For a three level model of NV$^-$, one would expect the saturation power to be inversely proportional to the absorption cross-section of NV$^-$.
However, this is not the case here, because of the ionization processes, as shown in Appendix~\ref{app:saturation}.
This is why the wavelength dependence of the saturation power of NV$^-$ does not resemble the inverse of the fluorescence of NV$^-$ shown in Fig.~\ref{fig:Fig_Ion_rates_wavelength}(c) measured at low illumination power (which is proportional to the NV$^-$ absorption cross-section).
\\
\section{Ionization and Recombination Energy}\label{sec:ion_energy}
\indent
In the following, we will determine the ionization and recombination energy of NV$^-$ in the diamond band gap.
For the wavelengths used in Sec.~\ref{sec:ion_rates} (540 - 610~nm), the ionization and recombination rates depend quadratically on the illumination power, because of the underlying two-photon processes, see Fig.~\ref{fig:Fig_energy_levels}.
However, if the photon energy is higher than the total ionization or recombination energy, both processes are possible with a one photon process.
In this case, the corresponding rate has a linear dependence on the excitation power.
Therefore, we measure the charge state dynamics in dependence of the excitation power at 435~nm to 520~nm.
Because direct observation of the dynamics is not possible at these wavelengths, we use correlated single shot charge state measurements.
\\
\indent
\begin{figure}
	\centering
		\includegraphics[width=0.49\textwidth]{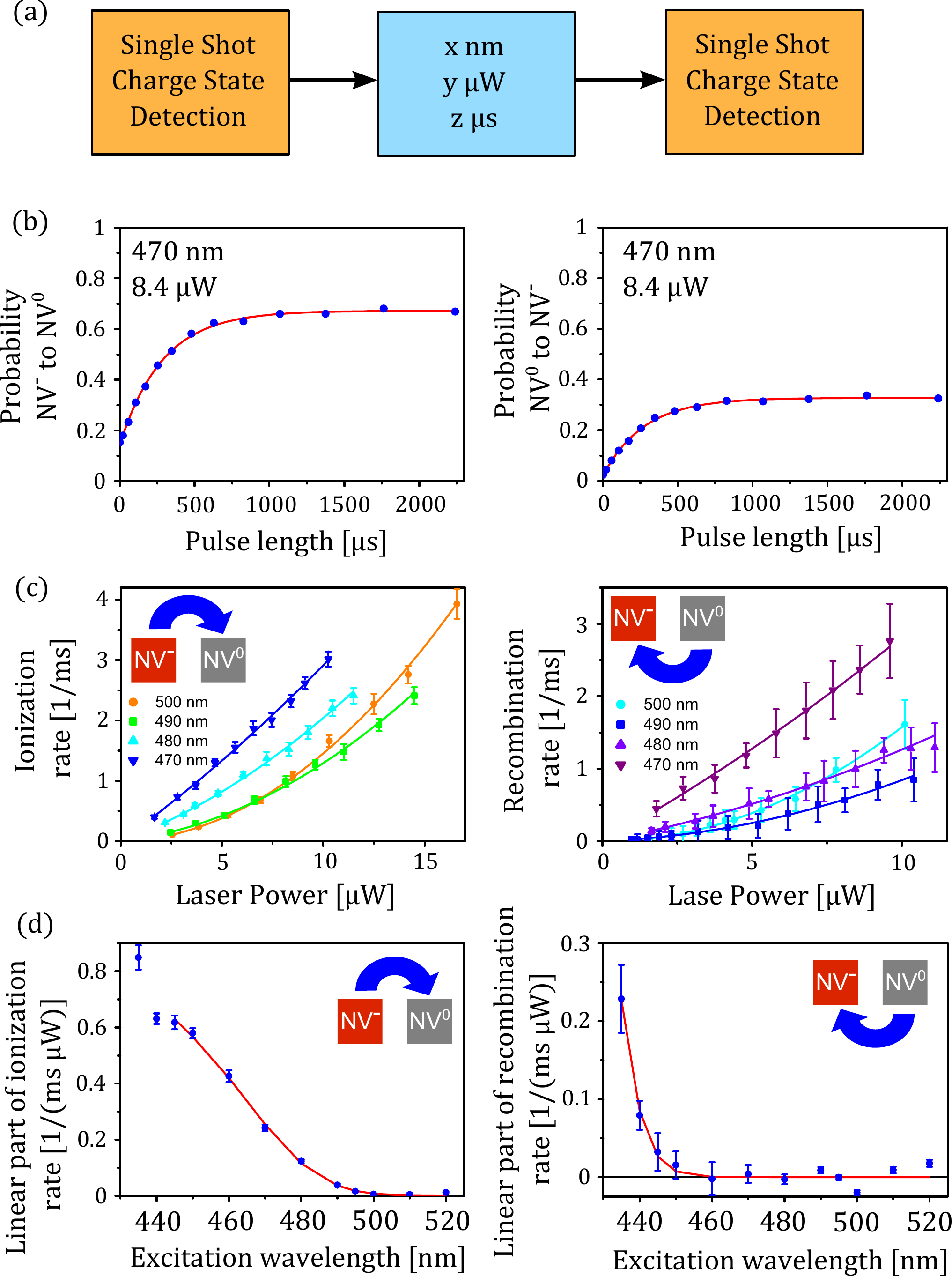}
	\caption{a) Pulse sequence for measuring the ionization and recombination rates for arbitrary illumination, by correlating the charge state before and after the laser pulse.
	b) Probability to change the charge state with the laser pulse in dependence of the pulse duration, for 470~nm and 8.4~$\mu$W, for initial states NV$^-$ and NV$^0$.
	The red line is an exponential fit.
	c) Ionization and recombination rates in dependence of illumination power $p$.
	The solid lines are parabola fits ($f(p) = ap+bp^2$).
	The error bars arise from statistical fit errors of the measurements shown in b).
	d) Linear part of the ionization and recombination rates (parameter $a$ of the above fit function) in dependence of the illumination wavelength.
	The red line is a fit as described in the Appendix.
	The error bars are are statistical fit errors of c).}
	\label{fig:Fig_Ionization_Energy}
\end{figure}
By setting a threshold in Fig.~\ref{fig:Fig_charge_state_detection}(b), we can determine the charge state of the NV non-destructively depending on whether the photon count number is above or below the threshold.
This allows us to do correlated charge state measurements, as shown in Fig.~\ref{fig:Fig_Ionization_Energy}(a).
After the first charge state detection, we know the current charge state of the NV.
Then, we apply an arbitrary laser pulse with certain wavelength and power, and determine the induced charge state dynamics by varying its duration.
Finally, a second single shot charge state measurement is applied.
By comparing the charge state before and after the arbitrary laser pulse, we get the probability that this laser pulse changed the NV's charge state.
Figure~\ref{fig:Fig_Ionization_Energy}(b) shows an example measurement result for 470~nm and 8.4~$\mu$W, in dependence of the pulse duration.
As shown in Appendix~\ref{app:ion_rate_equ}, we can calculate the ionization and recombination rates from this charge state dynamics for a specific excitation wavelength and power.
The power dependence of the rates is obtained by varying the power of the laser pulse, and measure the corresponding dynamics as shown in Fig.~\ref{fig:Fig_Ionization_Energy}(b).
\\
\indent
Figure~\ref{fig:Fig_Ionization_Energy}c) shows the rates in dependence of the illumination power for some wavelengths.
We see that for decreasing excitation wavelength, the curves become more linear.
To each curve, we fit the sum of a linear and of a quadratic function,
\begin{equation}
f(p) = ap + bp^2,
\end{equation}
where $p$ is the laser power.
The parameters $a$ and $b$ are the linear resp. quadratic part of the rates.
The linear part $a$ is plotted in Fig.~\ref{fig:Fig_Ionization_Energy}(d) in dependence of the illumination wavelength.
As explained in Appendix~\ref{app:ion_model}, we create a model for the ionization and recombination rates to account for thermal effects and the limited resolution of our supercontinuum laser.
This allows us to fit the measured data, and extract the ionization and recombination energy of NV$^-$.
With this model, we obtain an ionization energy of 2.60~eV, and a recombination energy of 2.94~eV.
\\
\section{Summary and Conclusion}
\indent
To summarize, we have performed detailed studies of the photo induced ionization and recombination process of NV$^-$ and NV$^0$.
We found that a recently reported "dark state" of NV$^-$ is indeed the neutral charge state NV$^0$ (Sec.~\ref{sec:dark_state}).
Real-time monitoring of the charge state dynamics via the NV's fluorescence was performed, which also yielded the bare excitation spectrum of NV$^-$ (Sec.~\ref{sec:ion_rates}).
We also determined the population of NV$^-$ and the fluorescence saturation behaviour for a broad range of the excitation wavelength (Sec.~\ref{sec:cds_saturation}).
The optimal excitation wavelength for NV$^-$ is found to be around 510 to 540~nm.
Finally, we determined the ionization energy of NV$^-$ to be 2.60~eV, and the recombination energy to be 2.94~eV (Sec.~\ref{sec:ion_energy}).
This was possible by observing the transition from a two-photon ionization and recombination process to a one-photon process, in dependence of the photon energy.
Note that the recombination energy of 2.94~eV is not the energy of the ground state of NV$^0$: The ground state of NV$^0$ can also be seen as a hole in the ground state of NV$^-$, and the photon has to provide the energy difference of the valence band and the ground state of NV$^-$ for the recombination.
This means that the ground state of NV$^-$ lies 2.6~eV below the conduction band and 2.94~eV above the valence band.
The sum of these two is 5.54~eV, close to the diamond band gap of 5.48~eV \cite{Clark_1964}.
\\
\indent
These results provide further insight into the properties and photo-physics of the NV.
A possible application of this process is stochastic high resolution microscopy, similar to PALM and STORM, which requires the activation of only few emitters in a diffraction limited area at the same time.
This is in principle possible with the NV, even in a reversible fashion, as can be seen in Fig.~\ref{fig:Fig_energy_levels}(b).
We expect that these results are not sensitive to the environment of the NV, but only depend on the band structure of the diamond.
Only if the inherent ionization / recombination rates (in dark) are comparable to the measurement time, e.g. through shifts of the Fermi level (cf. \cite{Grotz_NatureComm_2012}), an effect is expected.
A remaining question is the reason for the shape of the absorption spectrum of NV$^-$ shown in Fig.~\ref{fig:Fig_Ion_rates_wavelength}(c), and whether this two-maxima structure can be explained by the transition probabilities and density of states according to the vibrational modes in the ground and excited states, or if it arises from some other wavelength dependent mechanism.
\\
\indent
\begin{acknowledgments}
We thank P. Siyushev, T. Rendler, R. Koselov and S. Zaiser for experimental help and stimulating discussions.
Support by the Deutsche Forschungsgesellschaft (SFB/TR21, FOR 1482 and 1493), the European Commission (SOLID, DIAMANT and SQUTEC), the Volkswagenstiftung and the Landesstiftung BW is gratefully acknowledged.
\end{acknowledgments}
\appendix
\section{Saturation Behaviour}\label{app:saturation}
\begin{figure}
	\centering
		\includegraphics[width=0.4\textwidth]{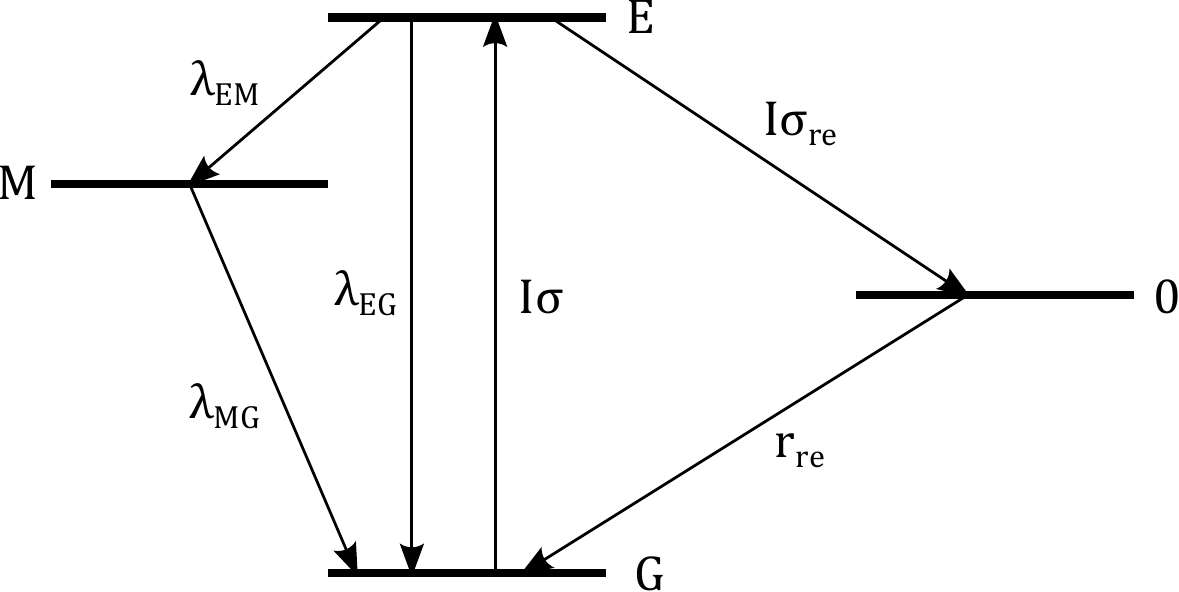}
	\caption{Illustration of the energy levels and rates, which are used to model the saturation behaviour.}
	\label{fig:Fig_app_levels}
\end{figure}

Here, we discuss the expected saturation behaviour of NV$^-$.
We use a four level model, which includes the NV$^-$ ground state (G), excited state (E) and metastable state (M), and one effective state for NV$^0$ (0), see Fig.~\ref{fig:Fig_app_levels}. 
We do not model the dynamics within NV$^0$, because we are only interested in the dynamics of NV$^-$.
The rate equations for these four states are
\begin{eqnarray}
\dot{p}_{\rm{G}} = -I\sigma p_{\rm{G}} + \lambda_{\rm{EG}}p_{\rm{E}} + \lambda_{\rm{MG}}p_{\rm{M}} + r_{\rm{re}}p_{\rm{0}}, \nonumber\\
\dot{p}_{\rm{E}} = I\sigma p_{\rm{G}} + (-\lambda_{\rm{EG}}-\lambda_{\rm{EM}}-I\sigma_{\rm{ion}})p_{\rm{E}}, \nonumber\\
\dot{p}_{\rm{M}} = \lambda_{\rm{EM}}p_{\rm{E}} - \lambda_{\rm{MG}}p_{\rm{M}}, \nonumber\\
\dot{p}_{\rm{0}} = I\sigma_{\rm{ion}}p_{\rm{E}} - r_{\rm{re}}p_{\rm{0}},
\end{eqnarray}
where $p_i$ denotes the population of state $i$, $\lambda_{ij}$ a decay rate from state $i$ to state $j$, $I$ the illumination intensity, $\sigma$ the absorption cross-section of NV$^-$, and $\sigma_{\rm{ion}}$ the absorption cross-section of NV$^-$.
The rate $r_{\rm{re}}$ is the effective recombination rate, which is
\begin{equation}
r_{\mathrm{re}} = \sigma_{\mathrm{re}}I\frac{I}{I+I_0},
\end{equation}
where $\sigma_{\mathrm{re}}$ is an effective recombination cross-section, and $I_0$ is the saturation power of NV$^0$.
This formula is has two components, the population of the NV$^0$ excited state, similar to eq. (\ref{eq:saturation}), multiplied by a term proportional to the illumination intensity $I$.
By setting all time derivatives in (A1) to zero,
\begin{equation}
0 = \dot{p}_{\rm{G}} = \dot{p}_{\rm{E}} = \dot{p}_{\rm{M}} = \dot{p}_{\rm{0}},
\end{equation}
and using
\begin{equation}
1 = p_{\rm{G}} + p_{\rm{E}} + p_{\rm{M}} + p_{\rm{0}},
\end{equation}
we can calculate the steady state populations of the four states.
We want to determine the expected fluorescence of the NV, in dependence of the illumination intensity.
The fluorescence of NV$^-$ is proportional to the population of the NV$^-$ excited state E, for illumination with moderate laser intensity, i.e. as long as the fluorescence decay rate $\lambda_{\rm{EG}}$ is much larger as the ionization rate $I\sigma_{\rm{ion}}$, which is the case for our measurements.
This population can be written as
\begin{equation}\label{eq:app_saturation}
p_{\rm{E}} = p_S\frac{I}{I + I_S},
\end{equation}
where the saturation population $p_S$ is
\begin{equation}
p_S = \frac{1}{1 + \lambda_{\rm{EM}}/\lambda_{\rm{MG}} + \sigma_{\rm{ion}}/\sigma_{rm{re}} + \sigma_{\rm{ion}}/\sigma},
\end{equation}
and the saturation power $I_S$ is
\begin{equation}\label{eq:app_sat_power}
I_S = \frac{\lambda_{\rm{EG}}+\lambda_{\rm{EM}}+I_0\sigma\sigma_{\rm{ion}}/\sigma_{\rm{re}}}{\sigma + \sigma\lambda_{\rm{EM}}/\lambda_{\rm{MG}} + 
	\sigma_{\rm{ion}} + \sigma\sigma_{\rm{ion}}/\sigma_{\rm{re}}}.
\end{equation}
Equation (\ref{eq:app_saturation}) is proportional to the fit function eq.~\ref{eq:saturation}, justifying its use.
Additionally, eq. (\ref{eq:app_sat_power}) shows that the saturation power is not inversely proportional to the absorption cross-section of NV$^-$, as would be the case without the ionization process.
Instead, it also depends on the ionization dynamics, and especially the saturation power $I_0$ of NV$^0$.
This is probably the reason for the steep increase of the saturation power in Fig.~\ref{fig:Fig_Saturation} at above 570~nm.
\\
\section{Ionization and recombination rate equations}\label{app:ion_rate_equ}
The charge state dynamics of the NV can be described by a system of two first order linear rate equations
\begin{equation}
\left(\begin{array}{c} \dot{p}_{\rm{NV}^-} \\ \dot{p}_{\rm{NV}^0} \end{array} \right) =
\underbrace{\left(\begin{array}{cc} - \lambda_{-0} & \lambda_{0-} \\ \lambda_{-0} & - \lambda_{0-} \end{array}\right)}_{A} \cdot
\underbrace{\left(\begin{array}{c} p_{\rm{NV}^-} \\ p_{\rm{NV}^0} \end{array}\right)}_{p},
\end{equation}
where $p_{NV^-}$, $p_{NV^0}$ are the population probabilities of NV$^-$ resp. NV$^0$, and $\lambda_{-0}$, $\lambda_{0-}$ are the rates from NV$^-$ to NV$^0$ resp. the other direction.
The time evolution of $p$ is then given by 
\begin{equation}
p(t) = e^{At}p(0),
\end{equation}
which can be solved by diagonalizing $A$.
In our case, the starting point $p(0)$ after the first charge state detection measurement is always either $\left(\begin{array}{c}1\\0\end{array}\right)$ (for NV$^-$) or $\left(\begin{array}{c}0\\1\end{array}\right)$ (for NV$^0$).
Then, the evolution is given by
\begin{equation}
p_{-/0}(t) = \left(\begin{array}{c} \frac{\lambda_{0-}}{\lambda_{-0}+\lambda_{0-}} \\ \frac{\lambda_{-0}}{\lambda_{-0}+\lambda_{0-}} \end{array}\right)
\pm \frac{\lambda_{0- / -0}}{\lambda_{-0}+\lambda_{0-}} e^{-(\lambda_{-0}+\lambda_{0-})t}\left(\begin{array}{c}-1\\1\end{array}\right),
\end{equation}
where $p_{-/0}(t)$ indicates the evolution after initializing in NV$^-$ resp. NV$^0$.
These two equations describe our measurement for the flip probabilities as shown in Fig.~\ref{fig:Fig_Ionization_Energy}(b).
From the steady state population $p(\infty)$ and the pumping rate $\lambda = \lambda_{-0}+\lambda_{0-}$ provided by fitting the measurement results yield the two rates $\lambda_{-0}$ and $\lambda_{0-}$, which are shown in Fig.~\ref{fig:Fig_Ionization_Energy}(c).
\\
\indent
\section{Model for measured ionization and recombination rates}\label{app:ion_model}
The basic idea is that due to thermal effects and due to the spectral width of our supercontinuum laser (of around 4~nm), we get a probability distribution for the energy absorbed by the electron (for NV$^-$) or by the hole (for NV$^0$).
Since we do not know the energy distribution of the NV$^-$ ground state due to thermal effects, we simply assume a Gaussian distribution $P(E_{\rm{f}})$ for the final energy $E_{\rm{f}}$ of the electron (hole) after absorbing a photon, to approximately account for both the thermal effects and the spectral width of the photon, i.e.
\begin{equation}
P(E_{\rm{f}}) \propto e^{-\frac{1}{2}\left(\frac{E_{\rm{f}}-\hbar\omega}{\sigma}\right)^2},
\end{equation}
where $\hbar\omega$ is the desired photon energy, and $\sigma$ is the width of the distribution, which will also be a fit parameter.
The total ionization rate is given by the convolution of the final energy of the electron (hole) with the density of states (DOS) for electrons in the conduction band (for holes in the valence band).
We approximate the band structure at the band edge with a parabola, such that the DOS $g$ is the same as for a free electron gas, which is
\begin{equation}
g(E) \propto \sqrt{E - E_0}, \qquad E \geq E_0, 
\end{equation}
where $E$ is the energy and $E_0$ the band edge energy relative to the ground state of NV$^-$.
$E_0$ is the parameter we would like to determine.
The rate therefore is
\begin{equation} 
r(\hbar\omega) = A\int^{\infty}_{-\infty} \sqrt{E_{\rm{f}} - E_0}\ e^{-\frac{1}{2}\left(\frac{E_{\rm{f}}-\hbar\omega}{\sigma}\right)^2} \rm{d} E_{\rm{f}}.
\end{equation}
This can be solved numerically and fit to the measurement data.
For the ionization rate, not all the data was used, because the approximation for the DOS in the band is only valid near the band edge.
Therefore, only the data up to 445~nm was used.
For the recombination rate, only four measurements points, which show a one photon process, are given.
Therefore, the dependency of the two fit parameters $E_0$ and $\sigma$ is very high.
For this fit, we used the same value for $\sigma = 0.069$~eV as obtained by the fit for the ionization rate.
This is also physically motivated, because the mechanism is the same, the spectral width of the photon and thermal effects.

\end{document}